\documentclass[conference, 10pt]{IEEEtran}

\usepackage{cite}

\ifCLASSINFOpdf
\else
\fi

\usepackage{amsmath}
\usepackage{upgreek}

\ifCLASSOPTIONcompsoc
 \usepackage[caption=false,font=normalsize,labelfont=sf,textfont=sf]{subfig}
\else
 \usepackage[caption=false,font=footnotesize]{subfig}
\fi

\usepackage{stfloats}

\usepackage[nolist]{acronym}
\usepackage{tikz}
\usepackage{wrapfig}
\usepackage{xcolor}[table]
\usepackage{colortbl}
\usepackage{pgfplots}
\usepackage{comment}
\usepackage{pgfplotstable}
\usepackage{booktabs}
\usepackage{caption}
\usepackage[inline]{enumitem}
\usepackage{multirow}
\usepackage{circuitikzgit}
\usepackage[Tol]{colorblind}
\usepackage{siunitx}
\usepackage{subcaption}

\usetikzlibrary{positioning}
\usetikzlibrary{shapes,arrows,patterns,calc,matrix,arrows.meta}

\IEEEoverridecommandlockouts
\IEEEpubid{\makebox[\columnwidth]{979-8-3195-0489-0/26/\$31.00~\copyright~2026 IEEE \hfill}\hspace{\columnsep}\makebox[\columnwidth]{ }}

\renewcommand\IEEEkeywordsname{Keywords}

\usepackage{eso-pic}

\newcommand\copyrighttext{%
  \footnotesize \copyright 2026 IEEE.  Personal use of this material is permitted.  Permission from IEEE must be obtained for all other uses, in any current or future media, including reprinting/republishing this material for advertising or promotional purposes, creating new collective works, for resale or redistribution to servers or lists, or reuse of any copyrighted component of this work in other works.}

\AddToShipoutPictureFG{%
  \put(\LenToUnit{0.5\paperwidth},\LenToUnit{40pt}){%
    \makebox(0,0){\fbox{\parbox{\dimexpr\textwidth-\fboxsep-\fboxrule\relax}{\copyrighttext}}}%
  }%
}

\begin{document}
\title{A Systolic Array Architecture for Nonlinear Activation Functions and Softmax Computation using Chebyshev Polynomials}

\author{\IEEEauthorblockN{Benedikt Schaible\IEEEauthorrefmark{1}, Anirudh Suresh Bharadwaj\IEEEauthorrefmark{2}, Ulf Schlichtmann\IEEEauthorrefmark{1} and Jiang Hu\IEEEauthorrefmark{2}\IEEEauthorrefmark{1}}
\IEEEauthorblockA{\IEEEauthorrefmark{1}Technical University of Munich}
\IEEEauthorblockA{\IEEEauthorrefmark{2}Texas A\&M University}}

\maketitle
\begin{acronym}
	\acro{ML}{Machine Learning}
	\acro{AI}{Artificial Intelligence}
	\acro{NN}{Neural Network}
	\acro{DNN}{Deep Neural Network}
	\acro{MLP}{Multilayer Perceptron}
	\acro{CNN}{Convolutional Neural Network}
	\acro{NAS}{Neural Architecture Search}
    \acro{PLA}{Piecewise Linear Approximation}
    \acro{RMSE}{Root Mean Square Error}
    \acro{AE}{Absolute Error}
    \acro{ARE}{Average Relative Error}
    \acro{ReLU}{Rectified Linear Unit}
    \acro{LLM}{Large Language Model}

	\acro{CPU}{Central Processing Unit}
	\acro{GPU}{Graphics Processing Unit}
	\acro{FPGA}{Field-Programmable Gate Array}
	\acro{CLB}{Configurable Logic Block}
	\acro{DSP}{Digital Signal Processor}
	\acro{RAM}{Random-Access Memory}
	\acroplural{RAM}[RAMs]{Random-Access Memories}
	\acro{BRAM}{Block Random-Access Memory}
	\acroplural{BRAM}[BRAMs]{Block Random-Access Memories}
	\acro{SoC}{System-on-Chip}
	\acroplural{SoC}[sections]{Systems-on-Chip}
	\acro{ASIC}{Application-Specific Integrated Circuit}
	\acro{PCIe}{Peripheral Component Interconnect Express}
	\acro{DMA}{Direct Memory Access}
	\acro{MAC}{Multiply-Accumulate}
    \acro{ROM}{Read-Only Memory}
    \acroplural{ROM}[ROMs]{Read-Only Memories}

    \acro{GEMM}{General Matrix Multiply}
	
	\acro{CU}{Compute Unit}
	\acro{PE}{Processing Element}
    \acro{LUT}{Lookup Table}
	\acro{NoC}{Network-on-Chip}
	\acroplural{NoC}[NoCs]{Networks-on-Chip}
    \acro{CORDIC}{Coordinate Rotation Digital Computer}
	
	\acro{HLS}{High-Level Synthesis}
	\acro{DSE}{Design Space Exploration}
	\acro{QoR}{Quality of Results}
	\acro{IP}{Intellectual Property}
	\acro{RTL}{Register-Transfer Level}
	\acro{HDL}{Hardware Description Language}
	\acro{VHDL}{Very High Speed Integrated Circuit Hardware Description Language}
	\acro{PnR}{Placement \& Routing}
	\acro{DSL}{Domain-Specific Language}
	\acro{I/O}{Input/Output}
	\acro{IR}{Intermediate Representation}
	\acro{MLIR}{Multi-Level Intermediate Representation}
	\acro{PPA}{Power, Performance, Area}

    \acro{KL}{Kullback-Leibler}
    \acro{RSE}{Row-Sum Error}
    
\end{acronym}

\setcounter{page}{1}
\IEEEpeerreviewmaketitle
\begin{abstract}
Neural Network Accelerators have gained popularity in recent years due to their greater efficiency than CPU-based platforms.
Often, these accelerators utilize different hardware units for univariate activation functions, such as tanh, and the multivariate softmax, thereby missing opportunities for resource sharing between them.
In this paper, we describe a novel systolic array-based activation unit architecture that supports multiple univariate activation functions as well as the softmax function.
By utilizing Chebyshev polynomial approximations, our activation function unit achieves up to 71\% lower mean absolute error for tanh compared to a \acs{CORDIC} baseline, while using 4.6\% less area and 5.1\% less power.
Our softmax approximation enables a 44.6\% and 79.0\% lower \acl{KL} divergence compared to \acs{CORDIC} and a piecewise-linear approximation, respectively.
\end{abstract}

\setlength{\textfloatsep}{10pt plus 1.0pt minus 1.0pt}
\setlength{\intextsep}{6pt plus 1.0pt minus 0pt}
\setlength{\floatsep}{0pt plus 2.0pt minus 0pt}

\section{Introduction}
\ac{ML} models such as \acp{NN} and transformers are commonly used across domains ranging from edge applications to large language models.
The widespread deployment of such models has led to growing demands for efficient computing platforms that can fulfill the requirements of the target application, such as power efficiency, throughput, or latency.

In many cases, custom hardware accelerators are capable of better fulfilling these design goals than off-the-shelf \ac{CPU} or \ac{GPU} based solutions due to their greater configurability~\cite{jouppi2017datacenter}.

When designing hardware accelerators, the computation of nonlinear functions is essential, as exact computation of these functions is often infeasible due to significant hardware requirements.
At the same time, the impact on accuracy needs to be kept at an acceptable level when using hardware-efficient approximations of nonlinear functions.
Specifically, two kinds of nonlinear functions need to be considered.
The first kind are the univariate activation functions, such as tanh and sigmoid, which are applied to the outputs of each neuron in a layer.

Using \ac{PLA} with few segments to compute univariate functions can achieve a reasonable approximation accuracy with relatively low resource requirements, while polynomial approximations can achieve higher accuracy at the cost of hardware complexity~\cite{andri2025flex}.
Another popular option is \ac{LUT}-based approximation, which can be sized to achieve the target accuracy~\cite{yang2018design}.
One disadvantage of \ac{LUT}-based methods is that resource sharing between activation functions is very limited, as each function needs its own \ac{LUT} mapping.

The second case is the multivariate softmax function, which is used, for instance, to obtain a probability distribution over the categories represented by the last layer of a \ac{NN}.
Efficient computation of softmax is also crucial in transformer workloads, as the softmax operation accounts for over a third of the runtime for large sequence lengths~\cite{stevens2021softermax}.
Additionally, low-accuracy approximations can lead to compounding errors with severe impact on the output accuracy for transformers.
\IEEEpubidadjcol

In contrast to univariate functions, softmax is not easily approximated directly by \ac{LUT}-based methods or \ac{PLA} due to its multidimensional nature.
Instead, \ac{PLA} or \acp{LUT} can be used to approximate the exponential function or division to obtain the output~\cite{vasyltsov2021efficient}.
Spagnolo et al.~\cite{spagnolo2021aggressive} implement softmax by utilizing a first-order Taylor approximation of $e^x$, which is then rounded to the nearest power of two to allow for bit shifts instead of division.
While their softmax approximation can identify the highest-probability class with high accuracy across their evaluated \acp{NN}, the overall accuracy of their softmax is significantly lower for complex networks when considering the top-3 classes.
Koca et al.~\cite{koca2023hardware} present an approximation for the softmax function that approximates division and $e^x$ in a way that allows both to be expressed in terms of addition, multiplication, and bit shifting.
Their approximation accuracy is relatively low, leading to a measurable impact when used in the BERT transformer~\cite{devlin2019bert}. 
Raghuram et al.~\cite{softmaxcodebase} utilize the \ac{CORDIC} algorithm for high-accuracy computation of softmax.
Their high-accuracy approximation comes at the cost of higher hardware requirements than \ac{PLA}-based methods.

In ONE-SA~\cite{sun2024onesa}, the authors add the capability of applying nonlinear activation functions, including softmax via \ac{PLA}, to their \acp{PE} in addition to the \ac{GEMM} operations that are usually carried out on the systolic array.
Although the paper enables more flexibility in supporting a wide range of activation functions, their description of the architecture leaves out details of how they implement the softmax function.

In this work, we present an architecture for computing both univariate activation functions and the softmax using a systolic array separate from the systolic array used for convolution and \ac{GEMM}.
Our architecture employs the capability of systolic arrays to calculate polynomial approximations of nonlinear functions.
We also use polynomials to approximate the softmax function, reusing the same hardware for both univariate activation functions and the multivariate softmax.
Our work makes the following contributions:
\begin{itemize}
    \item We propose a systolic array-based activation function unit architecture utilizing Horner's method~\cite{horner1819xxi} to compute approximations of nonlinear activation functions using Chebyshev polynomials~\cite{chebyshev1853theorie}.
    \item Our flexible architecture enables balancing approximation accuracy, level of parallelism, and hardware requirements according to the target application. 
    \item Our ASIC implementation supports the univariate tanh, sigmoid, and \ac{ReLU} functions as well as the multivariate softmax function and can easily be extended to implement other functions.
    \item Our approximation achieves 71\% and 41\% lower mean \acl{AE} compared to a feature-matched \ac{CORDIC} hardware design for tanh and sigmoid respectively, while requiring 4.6\% less area and 5.1\% less power.
    \item For large input dimensions ($N=256$), our softmax approximation outperforms both \ac{CORDIC} and ONE-SA with a 44.6\% and 79.0\% improvement respectively for the maximum \ac{KL} divergence.
\end{itemize}

\section{Mathematical Background}
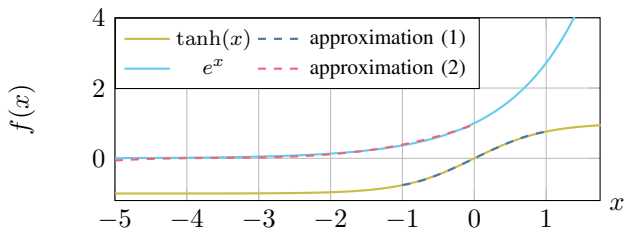
\begin{figure}[ht]
\centering
\begin{tikzpicture}
\begin{axis}[
    width=8cm,
    height=4cm,
    grid=both,
    xlabel={$x$},
    ylabel={$f(x)$},
    xmin=-5,
    xmax=1.75,
    ymin=-1.2,
    ymax=4,
    x label style={at={(axis description cs:1.0,0.2)},anchor=west},
    legend style={font=\footnotesize, at={(0,1)},anchor=north west,legend columns=2},
    domain=-5:2,
    samples=200 %
]
\addplot[T-Q-B4, thick] {tanh(x)}; %
\addlegendentry{$\tanh(x)$}

\addplot[T-Q-B1, dashed, thick, domain=-1:1] {-0.22*x^3 + 0.98*x}; %
\addlegendentry{approximation \eqref{eq:tanh-cheb}}
\addplot[T-Q-B2, thick]{e^x};
\addlegendentry{$e^{x}$}
\addplot[T-Q-B5, dashed, thick, domain=-5:0] {0.964 + 0.760*x + 0.206*x^2 + 0.019*x^3};
\addlegendentry{approximation \eqref{eq:exp-cheb}}
\end{axis}
\end{tikzpicture}
\vspace{-4pt}
\caption{Comparison of $\tanh(x)$ and $e^x$ and their polynomial approximations\label{fig:tanh-approx}.}
\vspace{-0.3cm}
\end{figure}

Chebyshev polynomials of the first kind are a polynomial sequence $T_n$. %
A linear combination of the $T_n$ polynomials, defined by $T_{n+1}(x)=2xT_n(x)-T_{n-1}(x)$ with $T_0(x)=1$ and $T_1(x)=x$, can be used to approximate continuous functions with low error~\cite{chebyshev1853theorie}.
The polynomial coefficients are obtained by interpolating the target function at Chebyshev nodes and converting the result into a standard power-series form (i.e., $a_n x^n + \dots + a_2 x^2 + a_1 x + a_0$)
Exemplarily, Fig.~\ref{fig:tanh-approx} shows polynomial approximations of $\tanh(x)$ for $x\in[-1,1]$ and $e^x$ for $x\in[-5,0]$, as defined in \eqref{eq:tanh-cheb} and \eqref{eq:exp-cheb} respectively.
\begin{gather}
    tanh(x) \approx -0.22x^3 + 0.98x, x \in [-1,1] \label{eq:tanh-cheb}\\
    e^x \approx 0.019x^3 + 0.206x^2 + 0.760x + 0.964, x \in [-5,0] \label{eq:exp-cheb}
\end{gather}
Particularly, the approximation of $\tanh(x)$ is a linear combination of $T_1(x)$ and $T_3(x)=4x^3-3x$ with a maximum approximation error of $0.0073$ in the interval $[-1,1]$.
\vspace{-4pt}

Horner's method~\cite{horner1819xxi} is the foundation for the efficient computation of these polynomials on a systolic array.
As shown in \eqref{eq:horner}, it relies on restating a polynomial as repeated multiplication of the previous intermediate result by $x$ and subsequent addition of the next polynomial coefficient $a_i$.
\begin{equation}
    a_3 x^3 + a_2 x^2 + a_1 x + a_0 = a_0 + x ( a_1 + x ( a_2 + x ( a_3 ))) \label{eq:horner}
\end{equation}

\section{Proposed Architecture}

\subsection{Systolic Array Architecture}

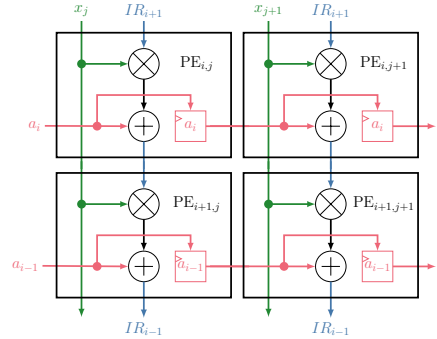
\begin{figure}[ht]
    \centering
    \resizebox{!}{0.19\textheight}{
    \begin{circuitikz}[]

        \begin{scope}
            \draw (0,0) node[mixer] (mul) {};
            \draw (0,-2) node[adder] (add) {};
            \draw[-latex, black, ultra thick] (mul) -- (add);

            \node (xin) at (-2,1.5) {};
            \node (xout) at (-2,-3.5) {};
            \node (xtext) at (-2,1.7) {\textcolor{T-Q-B3}{\Large$x_j$}};

            \draw (1.5,-2) node[registershape, t=\Large$a_i$, T-Q-B5] (areg) {};
            \node (ain) at (-3.5,-2) {\textcolor{T-Q-B5}{\Large$a_i$}};
            \node (aint) at (1.5,-1) {};
            \node (aout) at (3.5,-2) {};

            \node (irin) at (0,1.5) {};
            \node (irout) at (0,-3.5) {};
            \node (irtext) at (0,1.7) {\textcolor{T-Q-B1}{\Large$IR_{i+1}$}};

            \draw[-, T-Q-B3, ultra thick] (xin) -- (xout);
            \draw[*-latex, T-Q-B3, ultra thick] (-2.14,0) -- (mul.w);

            \draw[-latex, T-Q-B5, ultra thick] (ain.e) -- (add.w);
            \draw[*-latex, T-Q-B5, ultra thick] (-1.5,-2.14) |- (aint.center) -- (areg.n);
            \draw[-, T-Q-B5, ultra thick] (areg.e) -- (aout);

            \draw[-latex, T-Q-B1, ultra thick] (irin) -- (mul.n);
            \draw[-, T-Q-B1, ultra thick] (add.s) -- (irout);
            
            \draw[ultra thick] (-2.8,1) rectangle (2.8,-3);
            \node (PElabel) at (1.7,0) {\Large$\mathrm{PE}_{i,j}$};

        \end{scope}

        \begin{scope}[yshift=-4.5cm]
            \draw (0,0) node[mixer] (mul) {};
            \draw (0,-2) node[adder] (add) {};
            \draw[-latex, black, ultra thick] (mul) -- (add);

            \node (xin) at (-2,1.5) {};
            \node (xout) at (-2,-3.75) {};

            \draw (1.5,-2) node[registershape, t=\Large$a_{i-1}$, T-Q-B5] (areg) {};
            \node (ain) at (-3.8,-2) {\textcolor{T-Q-B5}{\Large$a_{i-1}$}};
            \node (aint) at (1.5,-1) {};
            \node (aout) at (3.5,-2) {};

            \node (irin) at (0,1.5) {};
            \node (irout) at (0,-4) {\textcolor{T-Q-B1}{\Large$IR_{i-1}$}};

            \draw[-latex, T-Q-B3, ultra thick] (xin) -- (xout);
            \draw[*-latex, T-Q-B3, ultra thick] (-2.14,0) -- (mul.w);

            \draw[-latex, T-Q-B5, ultra thick] (ain.e) -- (add.w);
            \draw[*-latex, T-Q-B5, ultra thick] (-1.5,-2.14) |- (aint.center) -- (areg.n);
            \draw[-, T-Q-B5, ultra thick] (areg.e) -- (aout);

            \draw[-latex, T-Q-B1, ultra thick] (irin) -- (mul.n);
            \draw[-latex, T-Q-B1, ultra thick] (add.s) -- (irout);
            
            \draw[ultra thick] (-2.8,1) rectangle (2.8,-3);
            \node (PElabel) at (1.7,0) {\Large$\mathrm{PE}_{i+1,j}$};

        \end{scope}

        \begin{scope}[xshift=6cm]
            \draw (0,0) node[mixer] (mul) {};
            \draw (0,-2) node[adder] (add) {};
            \draw[-latex, black, ultra thick] (mul) -- (add);

            \node (xin) at (-2,1.5) {};
            \node (xout) at (-2,-3.5) {};
            \node (xtext) at (-2,1.7) {\textcolor{T-Q-B3}{\Large$x_{j+1}$}};

            \draw (1.5,-2) node[registershape, t=\Large$a_i$, T-Q-B5] (areg) {};
            \node (ain) at (-4,-2) {};
            \node (aint) at (1.5,-1) {};
            \node (aout) at (3.5,-2) {};

            \node (irin) at (0,1.5) {};
            \node (irout) at (0,-3.5) {};
            \node (irtext) at (0,1.7) {\textcolor{T-Q-B1}{\Large$IR_{i+1}$}};

            \draw[-, T-Q-B3, ultra thick] (xin) -- (xout);
            \draw[*-latex, T-Q-B3, ultra thick] (-2.14,0) -- (mul.w);

            \draw[-latex, T-Q-B5, ultra thick] (ain.e) -- (add.w);
            \draw[*-latex, T-Q-B5, ultra thick] (-1.5,-2.14) |- (aint.center) -- (areg.n);
            \draw[-latex, T-Q-B5, ultra thick] (areg.e) -- (aout);

            \draw[-latex, T-Q-B1, ultra thick] (irin) -- (mul.n);
            \draw[-, T-Q-B1, ultra thick] (add.s) -- (irout);
            
            \draw[ultra thick] (-2.8,1) rectangle (2.8,-3);
            \node (PElabel) at (1.7,0) {\Large$\mathrm{PE}_{i,j+1}$};

        \end{scope}

        \begin{scope}[yshift=-4.5cm, xshift=6cm]
            \draw (0,0) node[mixer] (mul) {};
            \draw (0,-2) node[adder] (add) {};
            \draw[-latex, black, ultra thick] (mul) -- (add);

            \node (xin) at (-2,1.5) {};
            \node (xout) at (-2,-3.75) {};

            \draw (1.5,-2) node[registershape, t=\Large$a_{i-1}$, T-Q-B5] (areg) {};
            \node (ain) at (-4,-2) {};
            \node (aint) at (1.5,-1) {};
            \node (aout) at (3.5,-2) {};

            \node (irin) at (0,1.5) {};
            \node (irout) at (0,-4) {\textcolor{T-Q-B1}{\Large$IR_{i-1}$}};

            \draw[-latex, T-Q-B3, ultra thick] (xin) -- (xout);
            \draw[*-latex, T-Q-B3, ultra thick] (-2.14,0) -- (mul.w);

            \draw[-latex, T-Q-B5, ultra thick] (ain.e) -- (add.w);
            \draw[*-latex, T-Q-B5, ultra thick] (-1.5,-2.14) |- (aint.center) -- (areg.n);
            \draw[-latex, T-Q-B5, ultra thick] (areg.e) -- (aout);

            \draw[-latex, T-Q-B1, ultra thick] (irin) -- (mul.n);
            \draw[-latex, T-Q-B1, ultra thick] (add.s) -- (irout);
            
            \draw[ultra thick] (-2.8,1) rectangle (2.8,-3);
            \node (PElabel) at (1.7,0) {\Large$\mathrm{PE}_{i+1,j+1}$};

        \end{scope}

    \end{circuitikz}
    }
    \vspace{-4pt}
    \caption{Systolic array dataflow.}
    \label{fig:syst-arr-dataflow}
    \vspace{-0.4cm}
\end{figure}

Fig.~\ref{fig:syst-arr-dataflow} shows the dataflow through the systolic array on a $2 \times 2$ \ac{PE} section.
Each \ac{PE} receives the intermediate result $IR_{i+1}$ and the value $x$ from above and multiplies them.
Afterward, it adds the next polynomial coefficient $a_i$ and propagates its result $IR_i$ and $x$ to the \ac{PE} below, also forwarding $a_i$ to its right neighbor via a register.
The topmost \ac{PE} receives $0$ as its $IR$ input, so its output is $IR_n = a_n$, which is forwarded to the next \ac{PE}, continuing with the Horner computation until the $n+1$st \ac{PE} of the column outputs the final result of the degree $n$ polynomial evaluated at $x$.

Each column of $n+1$ \acp{PE} can evaluate polynomials up to degree $n$.
Polynomials with degree $d < n$ are computed by setting $a_i=0$ for $i>d$.

As all neurons of a single layer typically use the same activation function, all \acp{PE} in a row use the same polynomial coefficient $a_i$, such as $-0.22$ and $0.98$ in~\eqref{eq:tanh-cheb}.
Once the array starts computing a function, it takes one cycle per \ac{PE} column for the $a_i$ to propagate through the array, before the full throughput is reached.
Since \acp{NN} require very few activation functions, the coefficients of the approximation polynomials are implemented as hard-coded constants, with a multiplexer selecting the current activation function among the constants.

\vspace{-4pt}
\subsection{Activation Function-specific Aspects}\label{subsec:act-specific}

\subsubsection{Sigmoid and tanh}
For tanh and sigmoid, a single approximation polynomial, such as the one in \eqref{eq:tanh-cheb}, is computed using the systolic array.
The outputs of tanh and sigmoid are in the ranges $(-1,1)$ and $(0,1)$, respectively.
Both functions asymptotically approach their upper bound as $x$ goes to $+\infty$ and their lower bound as $x$ goes to $-\infty$.

To limit the input domain for which our approximation needs to be valid, we clip the output of tanh and sigmoid to their respective bounds outside of the valid approximation value range $[x_{min}, x_{max}]$.
Limiting the domain is desirable, as a larger domain leads to increased approximation error across the whole domain.
On the other hand, the closer $x_{min}$ and $x_{max}$ are to zero, the higher is the error from clipping directly outside of the interval.
Thus, when selecting the specific approximation polynomials, the values of $x_{min}$ and $x_{max}$ are important parameters that determine the tradeoff between the error inside and outside of $[x_{min},x_{max}]$.
The clipping means our approximations of tanh and sigmoid can be thought of as a piecewise function, as illustrated for tanh in~\eqref{eq:piecewise}.

\begin{equation}
    tanh(x) \approx \begin{cases}
        -1 & x < x_{min} \\
         \displaystyle \sum_{i=0}^{deg} a_ix^i& x_{min} \leq x \leq x_{max} \\
         1 & x > x_{max}
    \end{cases} \label{eq:piecewise}
\end{equation}

\subsubsection{\acs{ReLU}}\label{subsubsec:ReLU}
While the \ac{ReLU} function could be implemented as a clipped degree 1 polynomial, it can be trivially implemented by a dedicated \ac{ReLU} unit consisting of a comparator, enabling higher throughput with minimal additional hardware.
To maintain the same degree of parallelism as the systolic array, one \ac{ReLU} unit per \ac{PE} column is used.

\subsection{Softmax Computation}\label{subsec:softmax-algorithm}
While Chebyshev polynomials are well-suited for directly approximating univariate functions such as sigmoid and tanh, considerations must be made for the softmax function.
Instead of using a single polynomial to represent the entire function, only $e^x$ is approximated using a polynomial.
To limit the value range for which our approximation of $e^x$ needs to be valid, we use a modified version of the softmax that bounds the input values of the exponential function to be less than or equal to zero.
This is valid because shifting all softmax inputs by the same value leaves the output distribution unchanged: the resulting factor $e^{-c}$ appears in both the numerator and denominator and cancels.
Such shifted variants are also commonly used in other applications to prevent over- and underflows~\cite{Goodfellow-et-al-2016}. Our shifted softmax function $\mathrm{softmax}(x)$ is given in \eqref{eq:softmax}.

\begin{equation}
    \mathrm{softmax}(x)_i = \frac{e^{x_i-\mathrm{max}(x)}}{\sum^K_{j=1} e^{x_j-\mathrm{max}(x)}} \label{eq:softmax}
\end{equation}

The shifted softmax is calculated in 4 steps:
\begin{enumerate*}
    \item The input values are shifted by the maximum input value to $x_i-\mathrm{max}(x)$.
    \item For each $i$ of the $K$ vector elements, $e^{x_i-\mathrm{max}(x)}$ is calculated using a polynomial approximation for $e^x$ such as the one given in \eqref{eq:exp-cheb}. As new values arrive, they are each stored in an intermediate register and also summed up to $\sum^K_{j=1} e^{x_j-\mathrm{max}(x)}$.
    \item A hardware divider calculates ${1}\big/{\sum^K_{j=1} e^{x_j-\mathrm{max}(x)}}$.
    \item Finally, the result of step 3 is multiplied by the individual $e^{x_i-\mathrm{max}(x)}$ to obtain $\mathrm{softmax}(x)$.
\end{enumerate*}

As the intermediate values from step 2 need to be stored in intermediate registers, the maximum vector length $K$ is limited by the number of available registers.

Similar to tanh and sigmoid, our polynomial approximation of $e^x$ is clipped to zero below a certain $x_{min}$.
As all inputs of $e^x$ are at most zero, clipping above $x=0$ is not necessary.
This clipping removes low-probability inputs from the resulting probability distribution while keeping the inputs above the $x_{min}$ threshold.
For these non-clipped inputs, both their ranking and the ratios between their probabilities are preserved, while their absolute probabilities are scaled up by a common factor.
This shared factor arises from the clipped $e^x$ values that contribute zero to the softmax denominator.

\subsection{Selection of Approximation Polynomials}
\label{sec:selection-method}
The selection of the specific approximation polynomials needs to take several parameters into account.
Hardware designs for neural network accelerators often use fixed-point arithmetic to reduce hardware complexity compared to floating-point, which requires exponent handling and normalization.
Similar to integer representations, each bit in fixed-point representation represents a power of two, with the integer bits having values of $2^0$ and greater, while the fractional bits are those bits with value $2^{-1}$ and lower. 
We use the \textit{Qm.n} notation, describing an $(m+n+1)$-bit fixed-point number with $m$ integer bits, n fractional bits, and a sign bit.

When fixed-point arithmetic is used to compute the approximation polynomials, both the total bitwidth and the number of fractional and integer bits used to represent the polynomial coefficients play an important role.
The number of integer bits only needs to be sufficient to represent the polynomial coefficients.
Any further integer bits have no impact on approximation accuracy, unless a unified bitwidth across the whole accelerator is desired.
In that case, the value ranges of all inputs, outputs, and intermediate results must be considered to determine the required integer bits.
The number of available fractional bits influences the quantization error.
Thus, for any total bitwidth, it is preferable to have as many fractional bits as possible, while using as few integer bits as necessary.

In practice, a unified quantization format across all polynomial factors might be desirable to limit the hardware complexity.
Such a unified quantization would then need to use as many integer bits as necessary for the function with the highest integer-bit demand, sacrificing some fractional bit budget compared to per-function quantization schemes.

While the approximation error generally decreases with higher polynomial degrees in floating-point arithmetic, this correlation breaks down more quickly in hardware designs that use fixed-point representations.
One reason for this is that the quantization error of $x$ can be significantly amplified by high-order polynomials.
As a result, it is not optimal to choose the highest polynomial degree supported by the hardware to achieve optimum approximation accuracy.

As mentioned in Sec.~\ref{subsec:act-specific}, a third important factor is the value range $[x_{min},x_{max}]$, outside of which, the polynomial is clipped, balancing approximation and clipping error.

To identify the best candidate among all approximation polynomial candidates, all possible combinations across a range of these three parameters were generated and evaluated for their approximation accuracy.
The possible values for these parameters were
$\mathrm{degree} \in [1,12]$ and $\mathrm{total\ bitwidth} \in \{4,8,12,16\}$.
The value ranges for tanh and sigmoid were compared for both $[-2,2]$ and $[-3,3]$, while the exponential function used $[x_{min},0]$ with $x_{min} \in \{-2,-3,-4,-5\}$.

The tanh and sigmoid candidates were evaluated based on the \ac{RMSE}, mean \acf{AE}, and max.\ \ac{AE} across a value range of [-4,4] that includes the polynomial approximation and part of the clipped range.

As the exponential function is purely used to compute the softmax function, the top 100 candidates for $e^x$ based on their max.\ \ac{AE} on the interval $[-6,0]$ were then also ranked by their performance when used in softmax for a sample of 1000 vectors each for 8- and 32-input vectors.
To compare our approximate and the actual softmax function, the \ac{KL} divergence~\cite{kullback1951information}, a measure of difference between probability distributions, is used.
This was done to more accurately evaluate candidates for suitability for softmax, rather than for pure approximation error.

The selection process was performed using a two-step approach.
In the first step, each candidate polynomial was evaluated using the minimum number of integer bits required to represent its own coefficients and intermediate results, maximizing the fractional-bit budget per candidate.
From the top-ranked candidates of this first step across tanh, sigmoid, and $e^x$, we determined the maximum number of integer bits required by any of them.
In the second step, all candidates were re-evaluated under a unified quantization format using this fixed integer-bit count across all three functions, and the top-performing candidate of the second run for each function was selected as its final approximation.
If a non-unified quantization is preferred, only the first step is needed.
The resulting per-function quantization can use the same multiplier and adder hardware, but requires additional control logic to handle binary-point alignment depending on the per-function quantization.
The results of our two-step selection process, as well as the final choices of approximation polynomials for our designs, are discussed in Sec.~\ref{sec:eval-approximation}.

\subsection{Hardware Implementation}
We implement three hardware designs based on our proposed architecture, all using a systolic array with 10 rows and a signed 16-bit fixed-point representation, enabling the computation of degree 9 polynomials with 8 and 16 array columns, using the FreePDK45 process~\cite{stine2009freepdk}.
The designs differ in their parallel column count (8 or 16) and the number of intermediate value registers for softmax.

\section{Evaluation}

\subsection{Approximation Polynomials and Quantization} \label{sec:eval-approximation}
\input{figures/mae_tanh_4int}
\input{figures/mae_sigmoid_4int}
As described in Sec.~\ref{sec:selection-method}, a two-step approach was used to obtain the final approximation polynomials.
For both steps, the approximation accuracy was obtained in a quantization-aware manner.
While the arithmetic operations were carried out in floating-point accuracy, after every operation, the outputs were quantized to model the quantization impact.

The first step, where each candidate approximation uses the minimum number of integer bits required to represent its coefficients only, was used to determine the number of integer bits used in the final quantization.
In this step, for both tanh and sigmoid, the top candidate polynomials required only 2 integer bits, including the sign bit. So for these functions, a Q1.14 quantization would be optimal.
On the other hand, the candidate polynomials for the exponential function required more integer bits for their coefficients, with optimal quantizations ranging between Q2.13 and Q5.10 for the top 10 candidates.
Since we opted for a unified quantization scheme with a maximum bitwidth of 16 for all polynomial coefficients, each integer bit beyond Q1.14 to support a softmax approximation comes at the cost of a higher approximation error for tanh and sigmoid.
At the same time, the polynomial candidates approximating $e^x$ with fewer integer bits than three performed significantly worse than those with at least three integer bits, with diminishing returns beyond the third integer bit.
Thus, a unified quantization scheme of Q3.12 was determined to yield the best tradeoff between softmax and tanh/sigmoid accuracy.

Based on the decision to use a Q3.12 quantization scheme, the second polynomial selection step was performed with all coefficients using this quantization.
The mean \acp{AE} of tanh and sigmoid for 2001 evenly spaced points in the interval $[-4,4]$ are shown in Fig.~\ref{fig:mae_tanh_4int} and Fig.~\ref{fig:mae_sigmoid_4int} respectively.
Both figures are composed of two heatmaps, with subfigures (a) depicting the mean \ac{AE} for the polynomial candidates that are clipped outside of the interval $[-2,2]$ and subfigures (b) outside of the interval $[-3,3]$.
Each heatmap contains all bitwidth and polynomial combinations that were considered.

In general, the mean \ac{AE} decreases for higher bitwidths, while a higher polynomial degree does not always reduce the error.
As discussed in Sec.~\ref{sec:selection-method}, the approximation error initially improves for higher polynomial degrees, but at some point, the approximation error begins to increase again due to the higher impact of quantization error on high-degree terms.

The lowest mean approximation errors for a Q3.12 quantization were achieved for tanh with a degree 9 polynomial with clipping outside of $[-2,2]$, whereas the top sigmoid polynomial candidate is a degree 5 polynomial clipping outside of $[-3,3]$.
Both of these functions with minimum mean \ac{AE} also achieved the lowest maximum \ac{AE} in our further studies.
As for the \ac{RMSE}, the sigmoid candidate also has the lowest \ac{RMSE}, while the tanh candidate is only in the top-3, with the other two candidates with the same range and bitwidth, but degrees 8 and 7 placing in first and second respectively.

For the same quantization, the best-performing $e^x$ approximation has degree 3, with clipping below $-5$.
This polynomial has twice the max.\ \ac{AE} of another $e^x$ polynomial with the same range and degree 4, while still performing better in the \ac{KL} divergence than the degree 4 polynomial.

\subsection{Accuracy Comparison to other Works}
We compare the approximation accuracy of our designs to a \ac{CORDIC} implementation based on the hardware design by Raghuram et al.~\cite{softmaxcodebase}, which was extended to feature 8 parallel fixed-point \ac{CORDIC} units to match our systolic array's 8 parallel columns.

We also compare it with two piecewise linear approximation variants supported by ONE-SA~\cite{sun2024onesa} using segments of length 0.25 and 1, respectively, clipped outside of the same range $[x_{min},x_{max}]$ as our polynomial approximation.
Both designs use the same Q3.12 quantization.

Fig.~\ref{fig:activation-errors} shows the \ac{AE} of our approximation and the \ac{CORDIC} and ONE-SA comparisons for tanh and sigmoid, with the mean and max.\ \ac{AE} listed in Table~\ref{tbl:approximation-error}.

\pgfplotsset{
  afu error/.style={
    width  = \linewidth,
    height = 4cm,
    xlabel = {$x$},
    ylabel = {absolute error},
    ylabel near ticks,
    xlabel style={
      at={(0.975,0)},
      anchor=west,
    },
    scaled y ticks = base 10:2,
    ytick scale label code/.code={$\cdot 10^{-2}$},
    ymin= 0,
    grid   = major,
    grid style = {dashed, gray!40},
    legend style = {
      font        = \scriptsize,
      at          = {(0.45, 1.02)},
      anchor      = south,
      draw        = none,
      fill        = none,
      legend columns=-1,
    },
    legend cell align = left,
    tick label style  = {font=\small},
    label style       = {font=\small},
  },
}

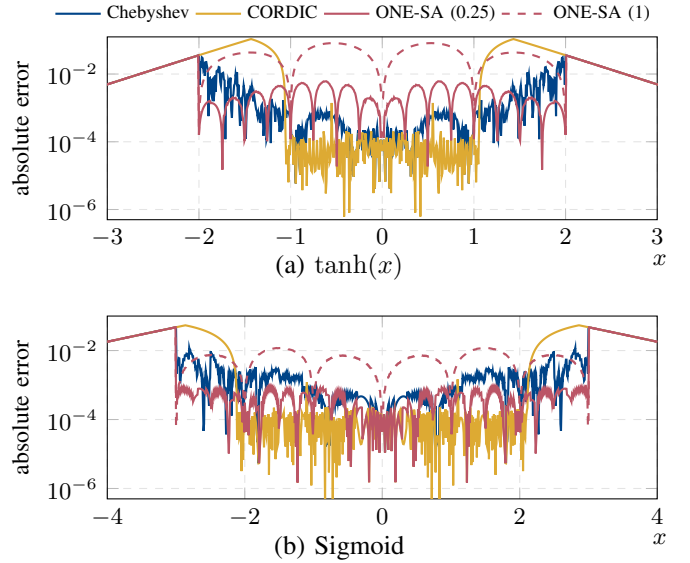
\begin{figure}[t]
  \centering

  \subfloat{
    \begin{tikzpicture}
      \begin{semilogyaxis}[
        afu error,
        xmin = -3, xmax = 3,
        ymin = 0.0000005, ymax = 0.125,
      ]
        \addplot+[mark=none, thick, T-Q-HC4] table [col sep=comma, x=x, y=err_cheb]{figures/errorplot-data/tanh.csv};
        \addlegendentry{Chebyshev}
        
        \addplot+[mark=none, thick, T-Q-HC2] table [col sep=comma, x=x, y=err_cordic_bp]{figures/errorplot-data/tanh.csv};
        \addlegendentry{CORDIC}
        
        \addplot+[mark=none, thick, T-Q-HC3] table [col sep=comma, x=x, y=err_pwl_025]{figures/errorplot-data/tanh.csv};
        \addlegendentry{ONE-SA (0.25)}

        \addplot+[mark=none, thick, T-Q-HC3, dashed] table [col sep=comma, x=x, y=err_pwl_10]{figures/errorplot-data/tanh.csv};
        \addlegendentry{ONE-SA (1)}

      \end{semilogyaxis}
      \node[anchor=south, yshift=-6pt] at (current bounding box.south) {(a) $\tanh(x)$};
    \end{tikzpicture}
    \label{fig:err-tanh}
  }
  
  \subfloat{
    \begin{tikzpicture}
      \begin{semilogyaxis}[
        afu error,
        xmin = -4, xmax = 4,
        ymin = 0.0000005, ymax = 0.1,
      ]
        \addplot+[mark=none, thick, T-Q-HC4] table [col sep=comma, x=x, y=err_cheb]
          {figures/errorplot-data/sigmoid.csv};
        \addplot+[mark=none, thick, T-Q-HC2] table [col sep=comma, x=x, y=err_cordic_bp]
          {figures/errorplot-data/sigmoid.csv};
        \addplot+[mark=none, thick, T-Q-HC3] table [col sep=comma, x=x, y=err_pwl_025]{figures/errorplot-data/sigmoid.csv};
        \addplot+[mark=none, thick, , T-Q-HC3, dashed] table [col sep=comma, x=x, y=err_pwl_10]
          {figures/errorplot-data/sigmoid.csv};
      \end{semilogyaxis}
      \node[anchor=south, yshift=-6pt] at (current bounding box.south) {(b) Sigmoid};
    \end{tikzpicture}
    \label{fig:err-sigmoid}
  }
  \caption{%
    \ac{AE} of fixed-point (Q3.12) activation function implementations, comparing our Chebyshev method to \acs{CORDIC} and ONE-SA using segment widths~$0.25$~and~$1$, respectively.%
  }
  \label{fig:activation-errors}
  \vspace{-0.3cm}
\end{figure}

\begin{table}[ht]
\centering
    \caption{\small Mean and Max.\ \ac{AE} for tanh and sigmoid on $[-4,4]$.}
    \label{tbl:approximation-error}
    \vspace{-6pt}
    \resizebox{\linewidth}{!}{
        \begin{tabular}{llSS}
        \toprule
        \textbf{Function} & \textbf{Method} & {\textbf{Mean \ac{AE}} $\mathbf{[}\cdot10^{-3}\mathbf{]}$} & {\textbf{Max \ac{AE}} $\mathbf{[}\cdot10^{-3}\mathbf{]}$} \\
        \addlinespace[-1pt]
        \midrule
        \multirow{4}{*}{tanh}    & Chebyshev     &  5.68 &  35.97 \\
                                 & \ac{CORDIC}   & 19.64  & 107.92  \\
                                 & ONE-SA (0.25) &  5.62 &  35.41 \\
                                 & ONE-SA (1.0)  & 24.79  &  82.00 \\
        \addlinespace[-2pt]
        \midrule
        \multirow{4}{*}{Sigmoid} & Chebyshev     &  8.95 &  47.43 \\
                                 & \ac{CORDIC}   & 15.23  &  53.96 \\
                                 & ONE-SA (0.25) &  7.88 &  47.07 \\
                                 & ONE-SA (1.0)  & 11.92  &  47.07 \\
        \addlinespace[-1pt]
        \bottomrule
        \end{tabular}
    }%
\end{table}

As shown, the 0.25 segment-width ONE-SA \ac{AE} is higher than ours for $|x| < 1.3$ for tanh, and slightly lower than ours for sigmoid, with the 1.0 segment-width version being significantly less accurate in both cases.
The \ac{CORDIC} approximation stops converging around $|x| > 1.1182$ for tanh~\cite{hu2002expandingCORDIC} and outside of $|x| > 2.236$ for sigmoid due to its reliance on \ac{CORDIC}'s tanh calculation, leading to a significantly higher error outside of the convergence ranges, resulting in a three times higher max.\ \ac{AE} for tanh.

As explained in Sec.~\ref{subsubsec:ReLU}, we compute ReLU using a comparator.
Similarly, ONE-SA can compute it exactly using two segments, resulting in no approximation error for both.

Table~\ref{tab:softmax-metrics} lists the  max.\ \ac{KL} divergence and \ac{RSE} observed across all samples drawn from a set of normal distributions $\mathcal{N}(0,\sigma^2)$ for $\sigma~\in~\{0.5,1,2,4\}$ and uniform distributions $U(-a,a)$ for $a \in \{0.5,1,2,4,6\}$ (128 samples per distribution) for our Chebyshev method, \ac{CORDIC}, and ONE-SA, for input vector sizes 8 and 256.
As ONE-SA~\cite{sun2024onesa} does not explicitly describe their softmax computation, we assume a calculation method similar to our 4-step approach from Sec.~\ref{subsec:softmax-algorithm}, but using a piecewise linear approximation of $\frac{1}{x}$ instead of a hardware divider.

Since the softmax output is a probability distribution, each output vector should sum to 1. 
Significant deviations from this may affect suitability for machine learning applications, where a row-sum of 1 is assumed.
All methods maintain low \ac{RSE} for vector size 8, with ONE-SA having the lowest \ac{KL} divergence, but the highest \ac{RSE}. Chebyshev outperforms both \ac{CORDIC} and ONE-SA for larger sequence lengths required in transformer and classification workloads with many classes, with substantially lower \ac{RSE} than both at $N=256$.
Our approach also has the lowest \ac{KL} divergence for large sequence lengths, demonstrating that its use of polynomial approximation and hardware division scales well for larger input vector sizes.
At larger $N$, ONE-SA's error stems from the division error, which scales with the input vector size.
For smaller input vector sizes, our approach falls behind ONE-SA due to its finer piecewise linear approximation of $e^x$.

\begin{figure*}[t]
  \centering
  \begin{minipage}{0.33\textwidth}
    \centering
      \footnotesize
       \captionof{table}{Max.\ \ac{KL} divergence and max.\ \ac{RSE} for softmax with vector size $N$. \label{tab:softmax-metrics}}
       \vspace{-6pt}
      \label{tbl:softmax-metrics}
      \resizebox{\linewidth}{!}{
      \begin{tabular}{l l r r}
        \toprule
        $\mathbf{N}$ & \textbf{Method} & \multicolumn{1}{c}{\textbf{Max KL}} & \multicolumn{1}{c}{\textbf{Max RSE}} \\
        \addlinespace[-1pt]
        \midrule
        \multirow{3}{*}{8} & Chebyshev & 0.393 & 0.007 \\
         & CORDIC & 3.006 & 0.002 \\
         & ONE-SA (0.25) & 0.035 & 0.015 \\
         \addlinespace[-2pt]
        \midrule
        \multirow{3}{*}{256} & Chebyshev & 3.252 & 0.021 \\
         & CORDIC & 5.879 & 0.061 \\
         & ONE-SA (0.25) & 15.512 & 1.021 \\
         \addlinespace[-1pt]
        \bottomrule
      \end{tabular}
      }%
  \end{minipage}
  \hfill
  \begin{minipage}{0.64\textwidth}
    \centering
    \centering
    \captionof{table}{Performance, power, and area of our designs (all with 10 \acp{PE} per column) and a \ac{CORDIC} activation function unit using FreePDK45~\cite{stine2009freepdk}.}
    \vspace{-6pt}
    \label{tbl:design-PPA}
    \resizebox{\linewidth}{!}{
    \begin{tabular}{lccccccccc}
    \toprule
    \textbf{Design} & \textbf{Parallel} & \textbf{Softmax} & \textbf{Area} & \textbf{Power} & \multicolumn{4}{c}{\textbf{Latency/output [ns]}} \\
    \addlinespace[-2pt]
    \cmidrule(lr){6-9}
    \addlinespace[-2pt]
     & \textbf{Factor} & \textbf{Registers} & [$\mathbf{\boldsymbol{\upmu} m^2}$] & [mW] & \textbf{tanh} & \textbf{Sigmoid} & \textbf{ReLU} & \textbf{Softmax} \\
    \addlinespace[-1pt]
    \midrule
    CORDIC     & 8  & 8   & 89719  & 0.8646 & 32.81 & 32.81 & ---  & 65.63 \\
    Our Design & 8  & 8   & 85634  & 0.8203 & 27.50 & 27.50 & 2.50 & 56.25 \\
    Our Design & 8  & 256 & 106192 & 0.9707 & 27.28 & 27.28 & 2.40 & 53.63 \\
    Our Design & 16 & 256 & 190242 & 1.7423 & 13.63 & 13.63 & 1.20 & 26.31 \\
    \addlinespace[-1pt]
    \bottomrule
    \end{tabular}
    }%
  \end{minipage}
\vspace{-0.3cm}
\end{figure*}

\begin{figure}[t]
\centering
\begin{tikzpicture}
\begin{axis}[
    width=\linewidth,
    height=4cm,
    ybar=2pt,
    bar width=7pt,
    enlarge x limits=0.15,
    ymin=0, ymax=2.8,
    ylabel={\hspace{-8pt}Ratio to CORDIC},
    symbolic x coords={Area, Power,{tanh\\latency}, {Sigmoid\\latency}, {Softmax\\latency}},
    xtick=data,
    xticklabel style={align=center},
    ytick={0, 0.5, 1.0, 1.5, 2.0},
    extra y ticks={1.0},
    extra y tick style={grid=major, grid style={dashed, gray!60}},
    extra y tick labels={},
    grid=both,
    grid style={dotted, gray!30},
    nodes near coords,
    nodes near coords style={
        font=\tiny, rotate=90, anchor=west,
        /pgf/number format/.cd, fixed, fixed zerofill, precision=2,
    },
    legend style={
        at={(0.42,1.02)}, anchor=south,
        legend columns=4, font=\small,
        /tikz/every even column/.append style={column sep=0.01cm},
    },
    legend cell align={left},
    cycle list={
        {fill=gray!70, draw=gray!70!black},
        {fill=blue!60, draw=blue!60!black},
        {fill=orange!70, draw=orange!70!black},
        {fill=green!60!black, draw=green!50!black},
    },
]

\addplot+[ybar] coordinates {
    (Area, 1.000) (Power, 1.000) ({tanh\\latency}, 1.000) ({Sigmoid\\latency}, 1.000) ({Softmax\\latency}, 1.000)
};
\addlegendentry{CORDIC (8/8)}

\addplot+[ybar] coordinates {
    (Area, 0.954) (Power, 0.949) ({tanh\\latency}, 0.838) ({Sigmoid\\latency}, 0.838) ({Softmax\\latency}, 0.857)
};
\addlegendentry{Ours (8/8)}

\addplot+[ybar] coordinates {
    (Area, 1.184) (Power, 1.123) ({tanh\\latency}, 0.831) ({Sigmoid\\latency}, 0.831) ({Softmax\\latency}, 0.817)
};
\addlegendentry{Ours (8/256)}

\addplot+[ybar] coordinates {
    (Area, 2.120) (Power, 2.015) ({tanh\\latency}, 0.415) ({Sigmoid\\latency}, 0.415) ({Softmax\\latency}, 0.401)
};
\addlegendentry{Ours (16/256)}

\end{axis}
\end{tikzpicture}
\caption{Area, power, and per-output latency of our designs normalized to the CORDIC baseline (lower is better). The legend format is (\emph{parallel factor} / \emph{Softmax register count}).}
\label{fig:design-PPA}
\vspace{-0.3cm}
\end{figure}
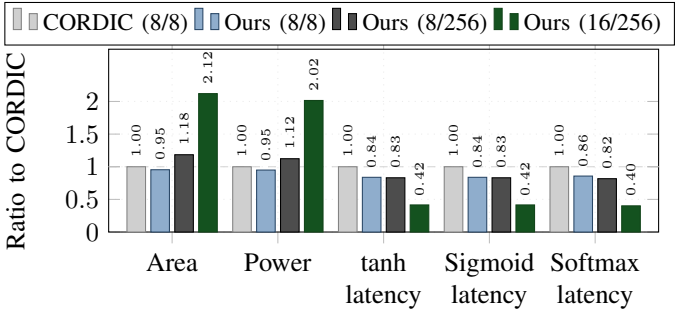

\subsection{Power, Area, and Latency}
Table~\ref{tbl:design-PPA} and Fig.~\ref{fig:design-PPA} show the area, power consumption, and latency of our designs and the \ac{CORDIC} baseline, synthesized with Synopsys Design Compiler V-2023 on the FreePDK45~\cite{stine2009freepdk} process.
The latency is given per output, assuming a filled pipeline.
As \ac{ReLU} can be implemented using a comparator, we exclude it from the \ac{CORDIC} baseline.
Our 8-parallel design with 8 softmax registers requires 4.6\% less area and 5.1\% less power than the \ac{CORDIC} baseline, while reducing tanh and sigmoid latency by 16.2\% and softmax latency by 14.3\%.
The 256-register variants trade higher area and power for the intermediate-value registers needed to support longer softmax vectors.
The 16-parallel 256-register variant uses twice as many columns as the 8-parallel variants but with the same intermediate-value register count, so its area is less than $2\times$ that of the 8-parallel 256-register design.

\section{Conclusion}
In this paper, we presented our systolic-array-based activation function unit that uses polynomial approximation to implement both the univariate \ac{ReLU}, tanh, and sigmoid functions, as well as the multivariate softmax function.
Our design achieves a substantially lower error than \ac{CORDIC} for tanh and sigmoid, while requiring less area and power.
Our approach also yields significantly lower \ac{KL} divergence than both \ac{CORDIC} and ONE-SA for large softmax dimensions.

\ifCLASSOPTIONpeerreview

\else
\section*{Acknowledgment}
With the support of the Technical University of Munich - Institute for Advanced Study, funded by the German Excellence Initiative, and NSF (CCF-2106725, CCF-2212346, CCF-2425399, CCF-2529764)
\fi

\bibliographystyle{IEEEtran}
\bibliography{library}

\end{document}